\documentclass[aps,twocolumn,superscriptaddress]{revtex4}
\usepackage{graphicx}
\usepackage{epsfig}
\usepackage{longtable}
\usepackage{amsmath}
\usepackage{amssymb}

\begin{document}

\title{Planar arrays of magnetic nanocrystals embedded in GaN}

\author{A.~Navarro-Quezada}
\email{andrea.navarro-quezada@jku.at}
\affiliation{Institut f\"ur Halbleiter- und Festk\"orperphysik, Johannes Kepler University, Altenbergerstr. 69, A-4040 Linz, Austria}

\author{T.~Devillers}
\email{thibaut.devillers@jku.at}
\affiliation{Institut f\"ur Halbleiter- und Festk\"orperphysik, Johannes Kepler University, Altenbergerstr. 69, A-4040 Linz, Austria}

\author{Tian Li}
\affiliation{Institut f\"ur Halbleiter- und Festk\"orperphysik, Johannes Kepler University, Altenbergerstr. 69, A-4040 Linz, Austria}

\author{A.~Bonanni}
\email{alberta.bonanni@jku.at}
\affiliation{Institut f\"ur Halbleiter- und Festk\"orperphysik, Johannes Kepler University, Altenbergerstr. 69, A-4040 Linz, Austria}

\date{\today}

\begin{abstract}
Single planar arrays of Ga$_x$Fe$_{4-x}$N magnetic nanocrystals embedded in GaN have been fabricated in an epitaxial process. The phase of the nanocrystals and their epitaxial relationship with the host matrix are studied $via$ high-resolution transmission electron microscopy and high-resolution x-ray diffraction. By changing the growth parameters and mode, the crystallographic phase and chemical composition of the nanocrystals can be varied on demand. In view of the different magnetic properties of the various phases, applications in room-temperature ferromagnetic as well as antiferromagnetic spintronic devices are envisaged.

\end{abstract}

\maketitle

Over the last decades, magnetic iron nitride (Fe$_{x}$N) compounds have attracted considerable attention owing to their striking magnetic properties, which make them suitable for applications in high density magnetic-recording write heads and media\,\cite{Jack:1951_PRSL, Kim:1972_APL, Coey:1999_JMMM, Takahashi:2000_JMMM}. 
We have recently demonstrated that either by co-doping with donors or acceptors, or by varying the growth temperature or the growth mode, we can control the aggregation of a variety of Fe$_{x}$N ferromagnetic (FM) and antiferromagnetic (AF) nanocrystals (NCs) in the magnetic condensed semiconductor (Ga,Fe)N grown on $c$-sapphire substrates\,\cite{Bonanni:2008_PRL,Rovezzi:2009_PRB,Navarro:2010_PRB, Navarro:2011_PRB}. In this way, this phase-separated material system is made suitable as building-block not only for FM- but also for AF-spintronic devices, these latter being an appealing perspective, due to the fact that a minimal critical current is needed to modify the magnetic order of an AF system\,\cite{MacDonald:2006_PRB} and since the resistance to external magnetic fields and the absence of stray fields make antiferromagnets particularly promising materials for ultrafast and ultrahigh-density spintronics\,\cite{Schick:2010_PRB}. 

The mentioned Fe$_x$N nanocrystals that aggregate in GaN doped with Fe above the solubility limit, have a narrow size distribution, which is independent of the phase and growth conditions. Furthermore, in contrast to the case of (Ga,Mn)As where a homogeneous alloy can be turned by post-growth annealing into a phase-separated system containing MnAs nanocrystals\,\cite{Hai:2011_JAP}, (Ga,Fe)N has been shown to be stable in the as-grown homogeneous or phase-separated phase under post-growth annealing up to 900\,$^\circ$C\,\cite{Navarro:2010_PRB}.

The most stable Fe$_x$N phase formed in phase-separated (Ga,Fe)N is wurtzite $\varepsilon$-Fe$_3$N, which is responsible for the robust FM character of the layers~\cite{Bonanni:2007_PRB, Bonanni:2008_PRL, Navarro:2010_PRB}. However, we observed the formation of another FM phase, namely $\gamma$'-Fe$_4$N~\cite{Navarro:2010_PRB}. This cubic phase forms at low temperatures and contains less nitrogen than its hexagonal counterpart $\varepsilon$-Fe$_3$N\,\cite{Jack:1952_AC}. Besides a high Curie temperature ($T_{\mathrm{C}}$) of 767\,K, this binary nitride possesses a remarkable chemical inertness and exhibits fascinating magnetic properties, such as the theoretically predicted~\cite{Kokado:2006_PRB}, and experimentally observed~\cite{Tsunoda:2011_APEX,Tsunoda:2009_APEX} highly enhanced negative spin polarization of electrons that makes it a suitable material for high-performance magnetic recording heads. Certainly, in order to exploit the outstanding properties of these magnetic nanocrystals in (Ga,Fe)N, and similar material systems~\cite{Jain:2011_JAP,Kuroda:2007_NM,Baik:2003_APL}, it is necessary to control their composition, size, magnetic properties and spatial distribution, to implement them in reliable device architectures, like $e.g.$ complex vertical heterostructures with spintronic functionalities~\cite{Narahara:2007_JJAP,Fushan:2008_APL,Jeff:2011_APL,Lee:2012_PRB}.

We have previously shown that the growth temperature does not only affect the crystallographic phase of the Fe$_x$N inclusions, but also their spatial distribution, and that the NCs tend to segregate towards the sample surface and in proximity of buried interfaces with $e.g.$ the nominally undoped high-quality GaN buffer grown between the sapphire substrate and the Fe-doped layer~\cite{Li:2008_JCG, Navarro:2010_PRB, Navarro:2011_PRB, Kowalik:2012_PRB}.\\

In this letter, we consider Ga$_x$Fe$_{4-x}$N nanocrystals with tunable composition determined by 0$<$$x$$<$1, embedded in a GaN host and we demonstrate that: (i) by varying the growth mode, we can control the spatial arrangement of the magnetic NCs into a $\textit{single}$ planar array perpendicular to the growth direction and located at a defined depth from the sample surface; (ii) the phase of the nanocrystals can be varied on demand $via$ the fabrication conditions; (iii) the NCs in the ensemble are ordered out-of-plane along the $c$-axis of the GaN matrix and in-plane in an orientation invariant for 30$^{\circ}$ rotations. We investigate the orientation and epitaxial relation of the NCs with respect to the GaN matrix and find that their lattice parameter increases with the Ga/(Fe+N) precursors ratio, implying a change in composition that is correlated to the magnetic properties of the NCs.\\

\begin{figure}[ht!]
	\begin{center}
	\includegraphics[width=\linewidth]{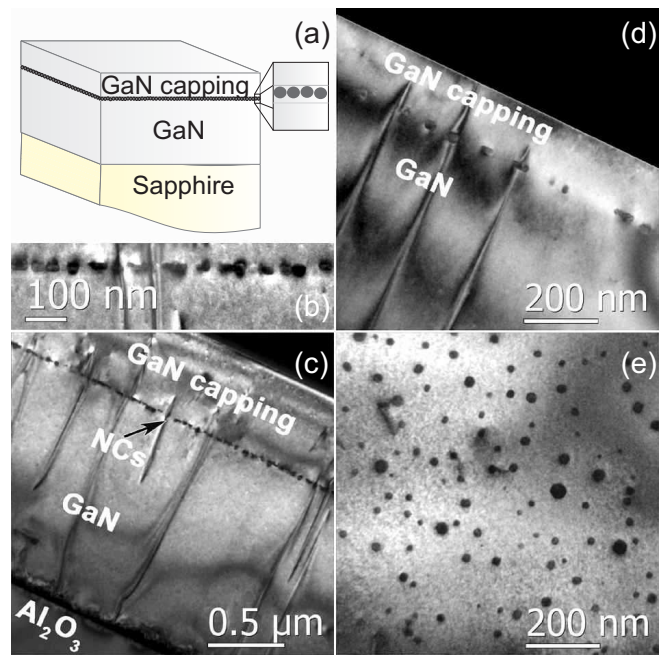}
	\caption{(a) Schematic sample structure; (b) cross-sectional TEM: magnification of the region containing the array of NCs for the sample reported in (c); (c) and (d) cross section TEM images of the samples, showing the spatial distribution of the nanocrystals into a planar array perpendicular to the growth direction and located 500\,nm and 150\,nm below the sample surface, respectively ; (e) plane-view TEM image of the sample in (c), giving the in-plane distribution of nanocrystals.}
	\label{fig:fig1}
	\end{center}
\end{figure}

The layers investigated in this work have been fabricated in a metal-organic-vapor-phase epitaxy (MOVPE) 200RF AIXTRON horizontal reactor system at a substrate temperature of 780\,$^\circ$C using trimethylgallium (TMGa), ammonia (NH$_3$) and ferrocene (Cp$_2$Fe) as precursors. The layers are grown on a 1\,$\mu$m high-quality GaN buffer layer, deposited on ${c}$-plane Al$_{2}$O$_{3}$ substrates, in a $\delta$-like growth mode obtained by alternately opening and closing the TMGa source while keeping the Cp$_2$Fe and NH$_3$ sources open. The NH$_3$ source flow is reduced from 1500 standard cubic centimers per minute (sccm) for the convenional GaN growth to 800 sccm during the growth of the $\delta$-layer, while the TMGa and Cp$_2$Fe source flows are varied from 1 to 5\,sccm and from 300 to 450\,sccm, respectively. We grow 15 $\delta$-periods, each consisting in the deposition of (Ga,Fe)N for 10 seconds and of FeN for 50 seconds, giving a total time of 1 minute/period. On top of the $\delta$-(Ga,Fe)N we deposit a GaN capping layer, whose thickness -- varied in the range between 150 and 500\,nm over the series of investigated samples -- determines the location of the array of nanocrystals below the sample surface. A schematic representation of the samples structure is provided in Fig.~\ref{fig:fig1}(a).\\
The arrangement of the NCs in a single planar array is made evident in the cross-section transmission electron microscopy (TEM) magnification reported in Fig.\,\ref{fig:fig1}(b) and in Fig.\,\ref{fig:fig1}(c) for a specimen with a 500\,nm and in Fig.~\ref{fig:fig1}(d) for one with a 150\,nm GaN capping layer. The TEM images are obtained from a JEOL 2011 Fast TEM microscope operating at 200\,kV, capable of an ultimate point-to-point resolution of 0.19\,nm permitting the imaging of lattice fringes with a 0.14\,nm resolution. The samples are prepared by standard mechanical polishing followed by Ar$^{+}$ ion milling at 4\,keV for about 1\,h.\\ 
As expected from the nominal growth sequence,the nanocrystals are located between the GaN buffer and the GaN capping. From the (HR)TEM analysis, segregation of the NCs at the interface between the substrate and the GaN buffer, as well as at the sample surface, is ruled out. The array of nanocrystals stabilizes at a distance from the sample surface that corresponds to the thickness of the capping layer, giving evidence that the aggregation of the NCs takes place during the $\delta$-(Ga,Fe)N growth. Moreover, the in-plane distribution of the nanocrystals is homogeneous, as seen in the plane-view TEM image reported in Fig.~\ref{fig:fig1}(e). The nanocrystals have a spheroidal shape with average size 21\,$\pm$\,6\,nm in-plane and 18\,$\pm$\,4\,nm in the growth direction, while the average sheet density in the plane is 9$\times$10$^{9}$\,$\pm$\,2\,NCs/cm$^{2}$. The density of nanocrystals in the array, can be adjusted by varying the Cp$_2$Fe source flow supplied during the $\delta$-layer growth.\\
From our previous studies on the growth of homogeneously doped (Ga,Fe)N layers, we know that the critical temperature for the formation of Fe-rich nanocrystals is 850\,$^\circ$C. Below this growth temperature no secondary phases have been observed for (Ga,Fe)N layers containing up to 1\% of Fe, neither by high-resolution x-ray diffraction (HRXRD), synchrotron-XRD nor HRTEM~\cite{Navarro:2010_PRB}. We can infer that the nanocrystal formation at this low deposition temperature is related to the $\delta$-layer growth mode, it takes place at the growing surface and is stopped as soon as the Cp$_2$Fe source flow is interrupted, since from energy dispersive electron spectroscopy we get no evidence of diffusion of Fe into the GaN capping nor into the underlying GaN buffer.\\
\begin{figure}[ht!]
	\begin{center}
	\includegraphics[width=\linewidth]{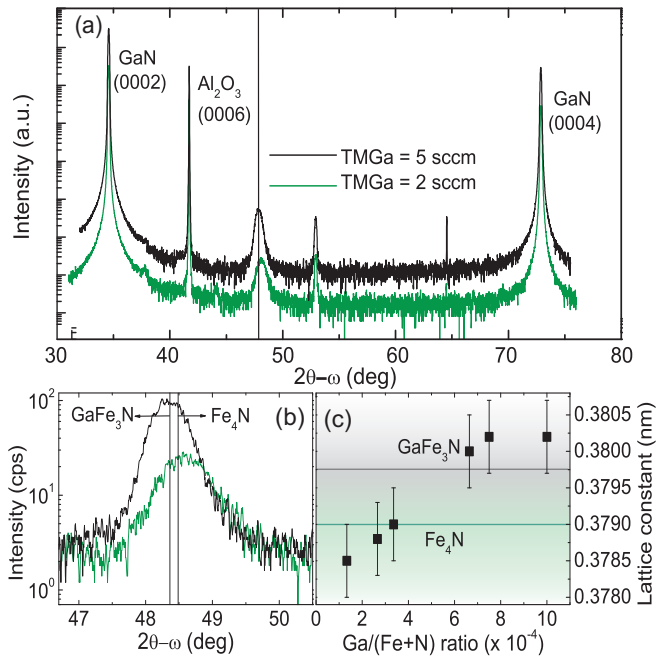}
	\caption{(Color online) (a) HRXRD spectra for two samples grown with different TMGa source flow showing the (200) diffraction peak of $\gamma$'-Ga$_x$Fe$_{4-x}$N. The solid line marks the theoretical position of $\gamma$'-GaFe$_3$N; (b) close-up of the (200) nanocrystal diffraction peak for the two samples showing a shift to higher angular values with decreasing TMGa; (c) evolution of the lattice parameter calculated from the diffraction peak position as a function of the overall Ga/(Fe+N) precursors ratio for different samples. The solid lines mark the theoretical values of the lattice parameter for $\gamma$'-GaFe$_3$N and $\gamma$'-Fe$_4$N.}
	\label{fig:fig2}
	\end{center}
\end{figure} 
\begin{figure*}
	\begin{center}
	\includegraphics[width=0.9\linewidth]{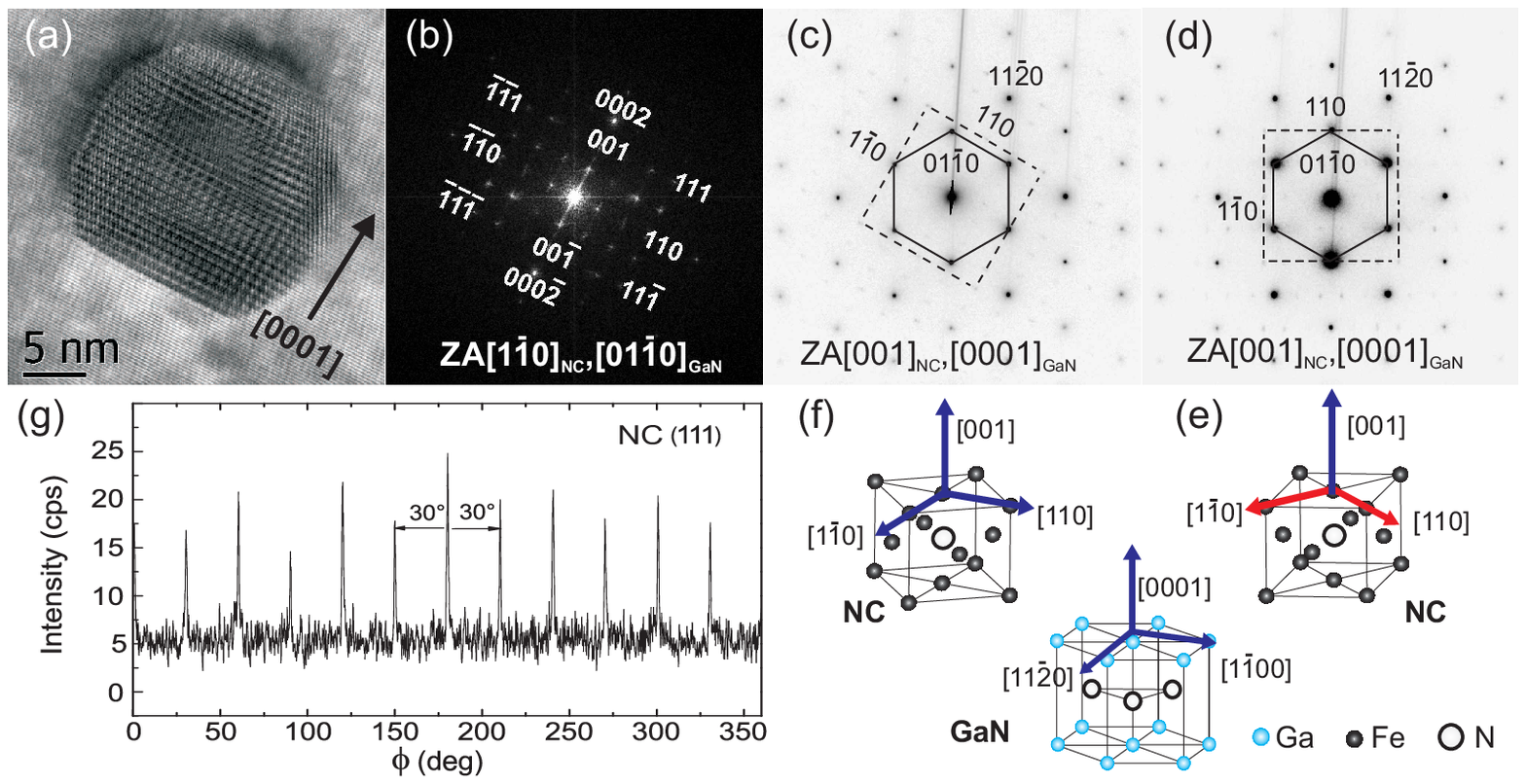}
	\caption{(Color online) HRTEM image of a nanocrystal viewed along the [01$\bar{1}$0] GaN zone axis; (b): diffractograms obtained by Fourier transformation of the HRTEM image in (a); (c) and (d): SADP of two NCs rotated in-plane by 30$^{\circ}$ about the (110) axis (and $c$-axis of the GaN host). Double-diffraction spots surround each of the primary reflections of GaN; (e) and (f): schematic drawing of the unit cell of the NCs in (d) and (c), respectively. The arrows indicate the orientation and rotation of the NCs with respect to the GaN host (middle sketch); (g): XRD angular $\phi$ scan around the (111) diffraction peak of the array of NCs, evidencing the in-plane 30$^{\circ}$ symmetry of the NCs.}
	\label{fig:fig3}
	\end{center}
\end{figure*}
The phase of the observed nanocrystals is obtained from HRXRD analysis. The measurements are carried out in an X'Pert Pro Material Research Diffractometer from Panalytical using an hybrid monochromator with a 1/4$^\circ$ divergence slit for the incoming beam and a PixCel detector with an active length of 0.5\,mm and a 5.0\,mm anti-scatter slit. Radial 2$\theta$-$\omega$ scans from 31$^\circ$ to 76$^\circ$ are acquired along the growth direction of GaN[0001], covering the range from the (0002) to the (0004) diffraction peaks. In Fig.\,\ref{fig:fig2}(a) the HRXRD spectra of two samples grown with different TMGa source flows are shown. The diffraction peak between 47.8$^\circ$ and 48.02$^\circ$, is identified as the (200) diffraction of fcc $\gamma$'-Ga$_x$Fe$_{4-x}$N with 0$<$$x$$<$1\,\cite{Rovezzi:2010_Report}. The average nanocrystal size along the growth direction is evaluated from the full-width-at-half-maximum (FWHM) of the (200) diffraction peak to be 15.5\,$\pm$\,0.6\,nm, consistent with the average size obtained from TEM. The lattice parameter of $\gamma$'-Ga$_x$Fe$_{4-x}$N varies linearly with the composition $x$ from 0.379\,nm ($x$\,=\,0) to 0.3797\,nm~($x$\,=\,1)\,\cite{Houben:2009_ChemMat}. From the obtained $d$-spacing for the Ga-rich sample (TMGa\,=\,5\,sccm) a lattice parameter of 0.3802\,$\pm$\,0.0005\,nm is calculated, matching the value for $\gamma$'-GaFe$_3$N, while for the Ga-poor samples the value shifts to 0.3788\,$\pm$\,0.0005\,nm, matching the one for $\gamma$'-Fe$_4$N. The shift in the diffraction peak position is clearly seen in Fig.\,\ref{fig:fig2}(b). In fact, we observe an increment in the lattice parameter with the overall Ga/(Fe+N) precursors ratio supplied during growth for different samples, as shown in Fig.\,\ref{fig:fig2}(c). 
Below a Ga/(Fe+N) precursors ratio of 5\,$\times$\,10$^{-4}$ the composition of the nanocrystals tends to the one of $\gamma$'-Fe$_4$N, while in a Ga-rich enviroment $\gamma$'-GaFe$_3$N is formed. Furthermore, the XRD spectra show that the nanocrystals have a preferential orientation [001]$_\mathrm{NC}$$||$[0001]$_\mathrm{GaN}$ with respect to the $c$-axis of GaN.\\ 

The in-plane orientation of the nanocrystals is obtained by analyzing both the selected area diffraction patterns (SADP) and the diffractograms of the HRTEM images on a statistically significant number of individual $\gamma$'-Ga$_{x}$Fe$_{4-x}$ nanocrystals. Based on the diffraction patterns, the interplanar spacings of the NCs are calculated using the DiffTools script in the \textit{DigitalMicrograph} software package\,\cite{Mitchell:2008_MRT}. 

As mentioned above, the orientation along the growth direction is common to all NCs. Furthermore, the analysis of the in-plane preferential orientations of the nanocrystals with respect to the GaN host reveals that the NCs order in equivalent in-plane orientations rotated by 30$^{\circ}$ about the (001) axis ($c$-axis of the GaN host). 
The HRTEM image of a $\gamma$'-Fe$_4$N nanocrystal taken along the [01$\bar{1}$0] GaN zone axis (ZA) in cross-sectional specimens, is reported in Fig.\,\ref{fig:fig3}(a). The diffractogram of the image in Fig.\,\ref{fig:fig3}(a), given in Fig.\,\ref{fig:fig3}(b), shows reflections originating from the superposition of spots from the NC (three indices) and from the GaN matrix (four indices), and the (001)[110]$_\mathrm{NC}$$||$(0001)[11$\bar{2}$0]$_\mathrm{GaN}$ epitaxial relationship is directly observed. Considering that no significant  distortion of the NC lattice is detected, according to the symmetries of the nanocrystals and of GaN, it can also be derived that: (001)[$\bar{1}$10]$_\mathrm{NC}$$||$(0001)[10$\bar{1}$0]$_\mathrm{GaN}$. This is confirmed by the SADP of a plan-view specimen reported in Fig.\,\ref{fig:fig3}(c), where the hexagon indicates the reflections from GaN, while the square indicates those from the NC, and the epitaxial relationship previously discussed is observed.
In the plan-view SADP of Fig.\,\ref{fig:fig3}(d), the in-plane [110]$_\mathrm{NC}$ is rotated by 30$^{\circ}$ with respect to the [01$\bar{1}$0] and [11$\bar{2}$0] directions of GaN. As an effect, the [$\bar{1}$10] overlaps with one of the six equivalent [11$\bar{2}$0] of GaN and the epitaxial relationship is preserved.  In fact, due to the symmetry of the NCs and of GaN, the in-plane rotation by 30$^{\circ}$ about the growth direction does not affect the epitaxial relationship. Double-diffraction spots are detected around each of the primary reflections in the SADPs of Fig.\,\ref{fig:fig3}(c) and (d). A sketch of the unit cell of the two representative NCs of Figs.\,\ref{fig:fig3}(c) and (d) and their orientation and rotation with respect to the $c$-axis of the GaN host, are provided in Figs.\,\ref{fig:fig3}(e) and (f). The XRD angular $\phi$ scan along the (111) diffraction peak of the NC in Fig.3(c), acquired in the range 0 to 360$^{\circ}$, confirms a 30$^{\circ}$ symmetry of the in-plane orientation of the NCs observed in HRTEM as an invariance upon 30$^{\circ}$ in-plane rotations of the in-plane orientation about the growth direction ($c$-axis of the GaN host).\\

In the ternary $\gamma$'-Ga$_x$Fe$_{4-x}$N alloy the magnetic moment increases with the amount of iron as we move closer to $x$\,=\,0\,\cite{Navarro:2010_PRB}. Thus, a systematic change from an AF to a FM ordering as a function of the composition takes place. While $\gamma$'-Fe$_4$N is FM with a $T_{\mathrm{C}}$ of 767\,K~\cite{Navarro:2010_PRB,Coey:1999_JMMM}, $\gamma$'-GaFe$_3$N exhibits a weak AF coupling with a Curie-Weiss temperature of -20 K~\cite{Houben:2009_ChemMat}. Furthermore, we have previously demonstrated \cite{Navarro:2010_PRB} that at growth temperatures above 850\,$^\circ$C, the phase $\zeta$-Fe$_2$N -- $\textit{antiferromagnetic}$ below 9\,K -- is stabilized. In this frame, for single-phase assemblies the magnetization of the nanocrystal arrays has a defined FM or AF character, depending on the composition of the nanocrystals. 
It is worth to recall here, that the Fe-rich NCs (with a phase dictated by the growth temperature and by the flow rates) are stabilized at the growth temperature and are stable upon post-growth annealing up to at least 900\,$^\circ$C~\cite{Navarro:2010_PRB,Navarro:2011_PRB}. In the case of arrays with a mixture of AF and FM NCs, on the other hand, due to the dominating FM character of the phases with $x$ close to 0, the $T_{\mathrm{C}}$ of the layers lies above room-temperature, as we previously observed in (Ga,Fe)N deposited at substrate temperatures above 900\,$^\circ$C~\cite{Navarro:2010_PRB,Navarro:2011_PRB}.\\

In summary, we have shown that the aggregation of $\textit{single}$ planar arrays of Fe-rich nanocrystals in GaN can be realized and the position of the single arrays in the sample volume and in a plane perpendicular to the growth axis can be adjusted on demand. The NCs have a narrow size distribution and an unambiguously defined epitaxial relation with the surrounding GaN matrix. The possibility to adjust the composition, phase, and consequently the magnetic features of these single arrays of NCs, makes these composite material systems suitable for the realization of complex spintronic devices like $e.g.$ non-volatile flash memories~\cite{Fushan:2008_APL,Jeff:2011_APL} and, remarkably, for FM- as well as for AF-spintronics.\\

\acknowledgments
This work has been supported by the European Research Council - ERC through the FunDMS Advanced Grant within the "Ideas" 7th Framework Programme of the European Comission and by the Austrian Fonds zur {F\"{o}rderung} der Wissenschaftlichen Forschung - FWF (P18942, P20065 and P22477).

\end{document}